\documentclass[usenatbib]{mn2e}
\usepackage{natbib}
\usepackage{apjfonts}
\usepackage{amsmath}

\usepackage{epsfig}
\usepackage{graphicx}
\usepackage{amssymb}
\usepackage{aas_macros}
\usepackage{color}

\newcommand{\kms}{\,{\rm km \, s^{-1}}}
\newcommand{\kpc}{\,{\rm kpc}}
\newcommand{\mpc}{\,{\rm Mpc}}
\newcommand{\oversim}[2]{\protect{\mbox{\lower0.5ex\vbox{%
   \baselineskip=0pt\lineskip=0.2ex
   \ialign{$\mathsurround=0pt #1\hfil##\hfil$\crcr#2\crcr\sim\crcr}}}}} 

\catcode`\"=\active\let"=\"  

\def\3{{\ss} }

\def\c12{{1\over 2}}

\def\d{{\rm d}}   
   
\def\plusplus{\raise 0.3ex\hbox{${\scriptstyle ++}$}{}}

 \def\r{\rightarrow} 
\def\and{{{\rm M}31}}
\def\mw{{\rm MW}}
\def\lmc{{\rm LMC}}
\def\smc{{\rm SMC}}
\def\lg{{\rm LG}}
\def\g{{\rm G}}
\def\gyr{{\rm Gyr}}
\bibliography{biblio}

\begin{document}   
\title[A timing constraint on the LMC mass]{A timing constraint on the (total) mass of the Large Magellanic Cloud}
\author[Jorge Pe\~{n}arrubia et al.]{Jorge Pe\~{n}arrubia$^{1}$\thanks{jorpega@roe.ac.uk}, Facundo A. G\'omez$^2$, Gurtina Besla$^3$, Denis Erkal$^4$ \& Yin-Zhe Ma$^5$\\
$^1$Institute for Astronomy, University of Edinburgh, Royal Observatory, Blackford Hill, Edinburgh EH9 3HJ, UK\\
$^2$Max-Planck-Institut fuer Astrophysik, Karl-Schwarzschild-Str. 1, D-85748, Garching, Germany\\
$^3$Department of Astronomy, University of Arizona, 933 North Cherry Avenue, Tucson, AZ 85721, USA\\
$^4$Institute of Astronomy, University of Cambridge, Madingley Road, Cambridge CB3 0HA, UK\\
$^5$Astrophysics and Cosmology Research Unit, School of Chemistry and Physics, University of KwaZulu-Natal, Durban, South Africa 
}
\maketitle  

\begin{abstract} 
This paper explores the effect of the LMC on the mass estimates obtained from the timing argument. 
We show that accounting for the presence of the LMC systematically lowers the Local Group mass ($M_\lg$) derived from the relative motion of the Milky Way--Andromeda pair. Motivated by this result we apply a Bayesian technique devised by Pe\~narrubia et al. (2014) to simultaneously fit (i) distances and velocities of galaxies within 3~Mpc and (ii) the relative motion between the Milky Way and Andromeda derived from HST observations, with the LMC mass ($M_\lmc$) as a free parameter. Our analysis returns a Local Group mass $M_\lg=2.64^{+0.42}_{-0.38}\times 10^{12}M_\odot$ at a 68\% confidence level. 
The masses of the Milky Way, $M_\mw=1.04_{-0.23}^{+0.26}\times 10^{12}M_\odot$, and Andromeda, $M_\and=1.33_{-0.33}^{+0.39}\times 10^{12}M_\odot$, are consistent with previous estimates that neglect the impact of the LMC on the observed Hubble flow. We find a (total) LMC mass $M_\lmc=0.25_{-0.08}^{+0.09}\times 10^{12}M_\odot$, which is indicative of an extended dark matter halo and supports the scenario where this galaxy is just past its first pericentric approach. Consequently, these results suggest that the LMC may induce significant perturbations on the Galactic potential.
\end{abstract}   


\section{Introduction} \label{sec:intro}
Despite being the two most luminous satellites of the Milky Way (MW hereinafter) the total masses of the Magellanic Clouds remain largely unconstrained. One of the key uncertainties is their accretion time. Given that satellite galaxies orbiting around a massive host lose a large fraction of their dark matter halo envelope to tides (e.g. Pe\~narrubia et al. 2008, 2009), an early accretion scenario would truncate the halo profile and limit the total mass to the mass enclosed within their luminous radius, i.e. $M\sim M(<r_{\rm max})= r_{\rm max}v_{\rm max}^2/G$, where $v_{\rm max}$ is the peak velocity of the circular velocity curve and $r_{\rm max}$ the outer-most radius with known members. With a circular velocity curve that peaks at $v_{\rm max}\simeq 90\kms$ and remains flat out to $8.7\kpc$ (van der Marel \& Kallivayalil 2014) the Large Magellanic Cloud (LMC hereinafter) has a (minimum) bound mass $M_\lmc(<8.7\kpc)= 1.7\times 10^{10}M_\odot$.
For the Small Magellanic Cloud (SMC hereinafter) $v_{\rm  max}\simeq 60\kms$ and $r_{\rm max}\simeq 3.5\kpc$ (Stanimirovi{\'c} et al. 2004), which yields $M_\smc(<3.5\kpc)= 2.9\times 10^{9}M_\odot$. Taking these estimates at face value and adopting the MW mass derived from the dynamics of galaxies in and around the Local Group ($M_\mw=0.8\times 10^{12}M_\odot$; Diaz et al. 2014; Pe\~narrubia et al. 2014; see also \S\ref{sec:dyn}) yields a mass ratio $(M_\lmc+M_\smc)/M_\g\simeq 0.02$,  
which suggests that the combined masses of the Magellanic Clouds do not contribute significantly to the mass budget of our Galaxy, $M_\g\approx M_\mw+M_\lmc+M_\smc$.

This result changes dramatically if the Magallanic Clouds are currently past their first pericentric approach to the MW, as suggested by current HST data (Besla et al. 2007; Kallivayalil et al. 2013). In the first-passage scenario the dark matter haloes of LMC and SMC may extend well beyond their luminous radius, such that $M\gg M(<r_{\rm max})$. How far beyond is currently an open issue given that the outer density profile and extent of dark matter haloes remain poorly constrained both theoretically and observationally. Lacking additional data one may invoke abundance-matching arguments, which suggest that galaxies with the luminosity of LMC ($M_\star\sim 2.9 \times 10^9 M_\odot$; van der Marel et al. 2002) reside in haloes with a virial mass $\sim 2\times 10^{11}M_\odot$ (Moster et al. 2013). Interestingly, $N$-body simulations of Besla et al. (2012) illustrate the SMC needs to complete 2--3 orbits about the LMC within the past $\sim 3\gyr$ in order to produce the main components of the Magellanic System which, 
given the high relative speed of the SMC with respect to the LMC ($\sim 130 \kms$; Kallivayalil 2013), requires $M_\lmc \gtrsim 10^{11} M_\odot$. 

Putting these estimates together and taking into account that the present-day Galactocentric distance of the LMC is $\simeq 50\kpc$ (McConnachie 2012)
yields a striking scenario where the LMC mass may be comparable to the mass of the MW at the LMC's location, $M_\mw(< 50\kpc)\approx 3\times 10^{11}M_\odot$ (Xue et al. 2008), and a virial radius $r_{\rm LMC, vir}\approx 110\kpc$ that is approximately twice the distance to the MW centre. Such a massive LMC would strongly perturb the Galactic potential and affect the constraints on the LG mass derived from the timing argument (G\'omez et al. 2015).


This Letter aims to constrain the total mass of the LMC by modelling the measured distances and velocities of galaxies in the Local Volume (hereinafter LV) as well as the relative motion between our Galaxy and M31 derived from HST observations (\S\ref{sec:frame}). 
Our analysis follows up on the work of Pe\~narrubia et al. (2014), who use the equations of motion that govern the dynamics of nearby galaxies in an expanding $\Lambda$CDM Universe to model 
how the Local Group (hereinafter LG) perturbs the local Hubble flow (\S\ref{sec:dyn}).
Model parameters are fitted using a Bayesian methodology outlined in \S\ref{sec:bayes}. Joint constraints on the parameters of interest are given in \S\ref{sec:results} and the results discussed in \S\ref{sec:sum}.

\section{The Local Group rest frame}\label{sec:frame}
We choose a coordinate system whose origin lies at the LG barycentre. Following Diaz et al.(2014, hereinafter D14) and Pe\~narrubia et al. (2014, hereinafter P14) the barycentre is found by demanding that the total momentum of our Galaxy (G) and Andromeda (M31) balances to zero. For simplicity we adopt the same vector notation as in D14, where ${\bf v}_{A\r B}$ denotes the velocity vector of $A$ measured from the reference frame $B$, such that ${\bf v}_{A\r B}= - {\bf v}_{B\r A}$ and ${\bf v}_{A\r B}={\bf v}_{A\r C} + {\bf v}_{C\r B}$.

Below we study the impact of the LMC on the (inferred) motion of the sun with respect to the LG barycentre. 
As a first step, let us calculate the total momentum of the LG as
\begin{eqnarray}\label{eq:angmom}
0&=&M_\g{\bf v}_{\g\r\lg}+ M_\and{\bf v}_{\and\r\lg}\\ \nonumber
&=&M_\lmc{\bf v}_{\lmc\r\lg} + M_\mw{\bf v}_{\mw\r\lg}+ M_{\and}{\bf v}_{\and\r\lg}.
\end{eqnarray}
For better comparison with previous work we introduce the parameters $f_m\equiv M_\g/M_\and$ and $f_c\equiv M_\lmc/M_\mw$, where $M_\g\simeq M_\lmc+M_\mw$. Using Equation~(\ref{eq:angmom}) it is relatively straightforward to show that the velocity vector of the sun with respect to the LG barycentre (the so-called solar {\it apex}) can be written as
\begin{eqnarray}\label{eq:apex}
{\bf v}_{\odot\r\lg}=-\frac{{\bf v}_{\and\r\mw}}{1+f_m}+{\bf v}_{\odot\r\mw}-\frac{f_m f_c}{(1+f_m)(1+f_c)}{\bf v}_{\lmc\r\mw},
\end{eqnarray}
where ${\bf v}_{\odot\r\mw}=(U_\odot,V_0+V_\odot, W_\odot)$ denotes the velocity vector of the sun with respect to the MW centre. We adopt a circular velocity of the MW at the solar radius $V_0=239\pm 5\kms$ (McMillan 2011), a Local Standard of Rest $(U_\odot,V_\odot, W_\odot)=(11.1, 12.2,7.2)\kms$ (Sch{\"o}nrich 2012) and a left-hand coordinate system ${\bf r}_{\odot\r\mw}=(-8,0,0)\kpc$. The velocity vectors of M31 and the LMC with respect to the MW centre are ${\bf v}_{\and\r\mw}={\bf v}_{\and\r\odot} + {\bf v}_{\odot\r\mw}$ and ${\bf v}_{\lmc\r\mw}={\bf v}_{\lmc\r\odot} + {\bf v}_{\odot\r\mw}$, respectively. Heliocentric velocity vectors are derived from line-of-sight velocities and proper motions. For Andromeda $(v_{\rm los},v_W,v_N)=(-300\pm 4,-119.2\pm 36,-61.9\pm 34)\kms$ (Sohn et al. 2012; D14), LMC $(262.2\pm 3.4,-465\pm 5,56\pm 11)\kms$ and SMC $(145.6\pm 0.6,-236\pm 19,-341\pm 19)\kms$ (McConnachie 2012; Kallivayalil et al. 2013).
 The heliocentric distances of M31, the LMC and SMC are $r_{\and\r\odot}=783\pm 25\kpc$, $r_{\lmc\r\odot}=51\pm 2\kpc$ and $r_{\smc\r\odot}=64\pm 4\kpc$, respectively (McConnachie 2012).

Unfortunately, we do not have access to the transverse motion of galaxies in the LV owing to their large heliocentric distances. Following Karachentsev 
\& Makarov (1996) one can derive a `corrected' radial velocity in the LG rest frame by adding the projection of the solar apex onto the heliocentric radial velocity 
\begin{equation}\label{eq:vrad}
v=v_{g\r\odot} + {\bf v}_{\odot\r\lg} \cdot \hat {\bf r}_{g\r\odot};
\end{equation}
where $v_{g\r\odot}$ denotes heliocentric line-of-sight velocities, and $\hat {\bf r}_{g\r\odot}\equiv {\bf r}_{g\r\odot}/|{\bf r}_{g\r\odot}|$ is a unit vector. At large distances Equation~(\ref{eq:vrad}) approximately corresponds to the radial component of the velocity vector, i.e. $v\approx {\bf v}_{g\r\lg}\cdot \hat{\bf r}_{g\r\lg}$.

Comparison between Equations~(\ref{eq:apex}) and~(\ref{eq:vrad}) shows that the presence of the LMC changes the barycentric velocity of nearby galaxies by $\Delta v_c=-f_m f_c(1+f_m)^{-1}(1+f_c)^{-1}({\bf v}_{\lmc\r\mw}\cdot \hat {\bf r}_{g\r\odot})$. This quantity can be positive or negative depending on the location of the galaxy with respect to the axis that joins the LMC with the MW centre, hence a wrong choice for $f_c$ will increase the scatter in the distance-velocity relation of galaxies randomly distributed on the sky. Even if the LMC contributes to a small fraction of the Galaxy mass ($f_c\ll 1$, $f_m\sim 1$), its large Galactocentric velocity, $v_{\lmc\r\mw}\simeq 327\kms$, leads to a non-negligible range $\Delta v_c\approx \pm  f_c 160 \kms $.
Accordingly, the maximum-likelihood value of $f_c$ is found in \S\ref{sec:bayes} by minimizing the scatter in the observed Hubble flow (see Appendix B of P14 for illustration).

It is straightforward to show that the mass of the SMC can be safely neglected in our analysis.
Application of the Tully-Fisher (1977) relation to the peak velocities of the SMC and LMC measured by Stanimirovi{\'c} et al. (2004) and van der Marel 
\& Kallivayalil (2014), respectively, yields $M_\smc/M_\lmc\sim (60\kms/90\kms)^4\sim 0.2 $. Using their measured separation, $|{\bf r}_{\smc\r\lmc}|\approx 24\kpc$, and relative velocity, $|{\bf v}_{\smc\r\lmc}|\approx 120\kms$, to balance the clouds momentum leads to a barycentre displaced by $\approx 4\kpc$ and $\approx 20\kms$ from the LMC, which can be neglected when compared with its current Galactocentric distance, $r_{\lmc\r\mw}\simeq 50\kpc$, and velocity, $v_{\lmc\r\mw}\simeq 327\kms$.
Using similar arguments van der Marel et al. (2012b) conclude that neither M33 nor M32 change the barycentre of M31 appreciably. 

\section{Timing argument \& the local Hubble flow}\label{sec:dyn}
In an expanding Universe the equations of motion that govern the dynamics of galaxies in the outskirts of the LG can be written as (Lynden-Bell 1981, Sandage 1986, Partridge et al. 2013, P14)
\begin{equation}\label{eq:kep}
{\ddot {\bf r}}=-\frac{G M}{r^2}\hat{\bf r} + H_0^2\Omega_{\Lambda}{\bf r},
\end{equation}
where $r=|{\bf r}_{g\r\lg}|$ is the distance of a galaxy to the LG barycentre, $M=M_\g + M_\and$ is the LG mass, $H_0=67 \pm 3\kms{\rm Mpc}^{-1}$ is the Hubble constant (Efstathiou 2013 and references therein) and $\Omega_\Lambda=0.686\pm 0.020$ is the dark energy density (Planck Collaboration 2013), which yields a Universe age $t_0\approx 13.7{\rm Gyr}$. 

For galaxies moving on radial orbits the solution to Equation~(\ref{eq:kep}) is univocally determined by measuring the barycentric distance ($r$) and velocity ($v$) at $t=t_0$. The isochrone $v=v(r,M,t_0)$ defined by Equation~(\ref{eq:kep}) is known as the `perturbed Hubble flow'. In this work isochrones are calculated under the assumption that (i) the LG potential is dominated by a static MW--M31 monopole, (ii) nearby galaxies can be treated as massless tracers and (iii) the LV is isolated. Tests in P14 show that both the quadrupole and the time-dependence of the LG potential have a negligible impact on the observed Hubble flow, while in \S\ref{sec:bayes} we use statistical methods to account for tangential motions that may arise from the self-gravity of the tracer population and the presence of galaxy associations in the vicinity of the LG. 

Neglecting the  energy term in Equation~(\ref{eq:kep}) and adopting $\d M/\d t=0$ yields a Keplerian solution to Equation~(\ref{eq:kep})
\begin{eqnarray}\label{eq:rtkep}
r&=&a(1-e\cos\eta) \\ \nonumber
t&=&\bigg(\frac{a^3}{G M}\bigg)^{1/2}(\eta - e\sin \eta),
\end{eqnarray}

where $\eta$ is the `eccentric anomaly', 
$a=L^2/[G M (1-e^2)]$ is the semi-major axis, $e=1+[2E L^2/(GM)^2]$ is the orbital eccentricity, $E=1/2v^2-GM/r$ and ${\bf L}={\bf r}\times {\bf v}$ are the specific energy and angular momentum, respectively. 
We consider orbits where $\eta=0$ at $t=0$; $\eta=\pi$ at turn-around radius, $r_{\rm apo}=a(1+e)$; and $\eta=2\pi$ at pericentre, $r_{\rm peri}=a(1-e)$. The radial and tangential velocity components are
\begin{eqnarray}\label{eq:vkep}
v_{\rm rad}&=& \bigg(\frac{a}{G M}\bigg)^{1/2}\frac{e\sin \eta}{1-e\cos\eta}\\ \nonumber
v_{\rm tan}&=& \bigg(\frac{a}{G M}\bigg)^{1/2}\frac{\sqrt{1-e^2}}{1-e\cos\eta}.
\end{eqnarray}

Equations~(\ref{eq:rtkep}) and~(\ref{eq:vkep}) can be used to measure the combined mass of our Galaxy and Andromeda, $M=M_\and+M_\g$, the so-called `timing argument' (e.g. van der Marel 
\& Guhathakurta 2008 and references therein). Position and velocity vectors now describe the relative motion between M31 and our Galaxy, i.e. $r=|{\bf r}_{\and\r\g}|$, $v_{\rm rad}={\bf v}_{\and\r\g}\cdot \hat {\bf r}_{\and\r\g}$ and $v_{\rm tan}=|{\bf v}_{\and\r\g}\times \hat {\bf r}_{\and\r\g}|$, where $\hat {\bf r}_{\and\r\g}$ is a unit vector pointing toward M31 from the Galaxy barycentre. Following similar steps as in \S\ref{sec:frame} we find
\begin{eqnarray}\label{eq:vm31g}
{\bf v}_{\and\r\g}={\bf v}_{\and\r\mw}-\frac{f_c}{1+f_c}{\bf v}_{\lmc\r\mw},
\end{eqnarray}
and similarly for ${\bf r}_{\and\r\g}$. 

\begin{figure}  
\begin{center}
\includegraphics[width=82mm]{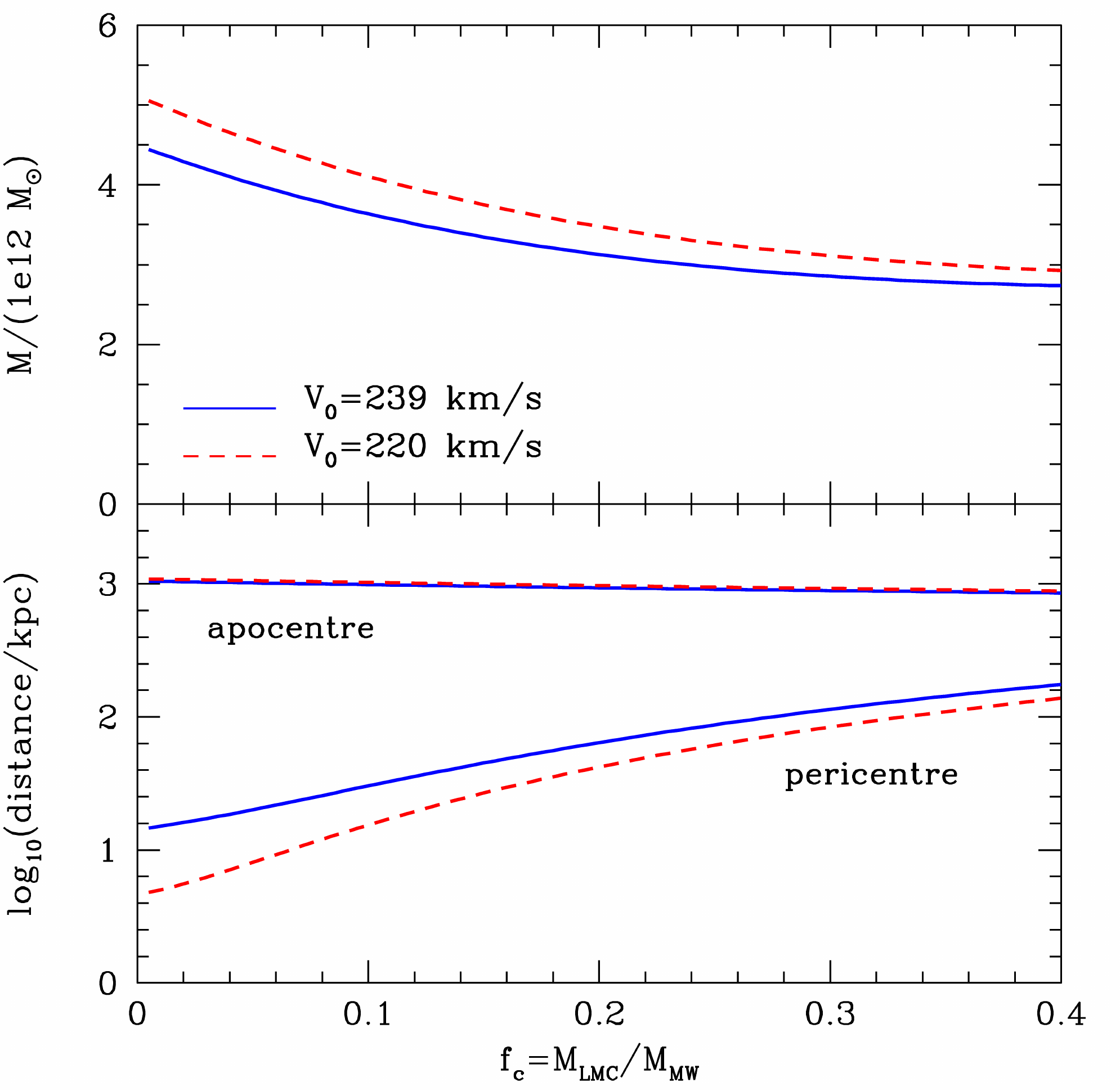}
\end{center}
\caption{LG mass (upper panel), and peri- and apo-centre distances between our Galaxy and M31 (lower panel) derived from the {\it timing argument}, Equations~(\ref{eq:rtkep}) and~(\ref{eq:vkep}), as a function of the mass ratio $f_c=M_\lmc/M_\mw$ for two values of the MW's circular velocity at the solar radius ($V_0$). Note that the LG mass $M=M_\and + M_\g$ and the impact parameter of the future Galaxy-M31 encounter strongly depend on the value of $f_c$.}
\label{fig:timing}
\end{figure}

Fig.~\ref{fig:timing} shows that the LG mass (upper panel) and the orbital parameters of the Galaxy--M31 pair (bottom panel) depend on the {\it unknown} value of $f_c=M_\lmc/M_\mw$. 
Models that neglect the LMC mass, $f_c \approx 0$, recover the main results found by van der Marel 
\& Guhathakurta (2012a), that is a LG mass, $M\approx 4.5\times 10^{12}M_\odot$, and an orbital pericentre $r_{\rm peri}\lesssim 20\kpc$, which imply that our Galaxy and M31 will suffer a head-on encounter $\sim 4\gyr$ from now (van der Marel et al. 2012b). As discussed in P14, the mass estimate increases up to $M\simeq 5.2\times 10^{12}M_\odot$ if one adopts 
$V_0=220\kms$, in agreement with Li \& White (2008). 

These estimates change significantly if the mass of the LMC exceeds the mass ratio derived from its internal kinematics.
In particular, for $f_c\gtrsim 0.2$ the LG mass drops down to $M\sim 2.8\times 10^{12}M_\odot$, in better agreement with the values independently derived by D14, $M=(2.5\pm 0.4)\times 10^{12}M_\odot$, and P14, $M=(2.3\pm 0.7)\times 10^{12}M_\odot$, while the pericentric distance increases up to $r_{\rm peri}\sim 100\kpc$, indicating a fly-by encounter rather than a head-on collision. Excluding models where the apocentric distance (i.e the turn-around radius) is smaller than the current separation between our Galaxy and M31, i.e. $r_{\rm apo}\gtrsim 800\kpc$, imposes an upper bound $f_c\lesssim 0.5$. It is thus clear that a robust measurement of the LG mass using the timing argument requires a tight constraint on the (total) LMC mass. Below we attempt to measure $f_c$ by modelling the dynamics of LV galaxies.

\begin{figure*}  
\begin{center}
\includegraphics[width=169mm]{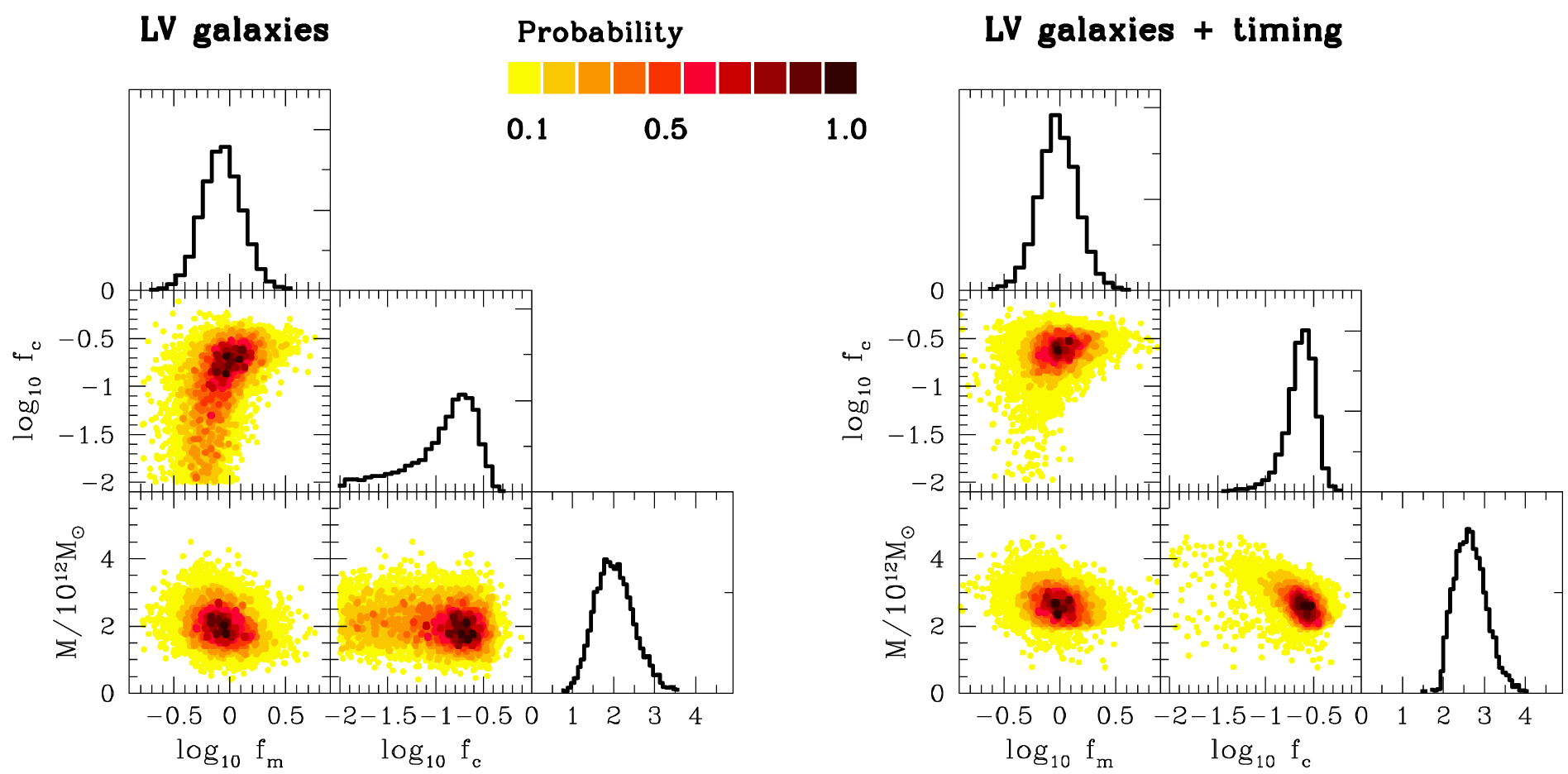}
\end{center}
\caption{ Sampling of posterior distributions of the LG mass, $M$, the mass ratios between the Galaxy and M31, $f_m=M_\g/M_\and$ and the LMC and the MW, $f_c=M_\lmc/M_\mw$. These distributions marginalize over $H_0$, $\Omega_\Lambda$, $V_0$ and $\sigma_m$ (see text). Left panels use the likelihood~(\ref{eq:likelv}) to fit the dynamics of
galaxies with $0.78 \lesssim r/\mpc \lesssim 3.0$. Right panels apply a weighted joint likelihood~(\ref{eq:likelg}) 
to fit the local Hubble flow and the motion of the Galaxy--M31 pair simultaneously.}
\label{fig:grid}
\end{figure*}
\section{Bayesian inference}\label{sec:bayes}
The main goal of this paper is to constrain the individual masses of the LMC, the MW and M31 from the measured locations and velocities of nearby galaxies taking into account statistical uncertainties in the circular velocity of the sun around the MW centre, $V_0$, as well as in the cosmological parameters $H_0$ and $\Omega_\Lambda$. 
To this end we use the following likelihood function (see P14 for details)
\begin{equation}\label{eq:likelv}
\mathcal{L}_{\rm LV}(\{D_i,l_i,b_i,V_{h,i}\}^{N_g}_{i=1}|\vec{S})=\prod_{i=1}^{N_g}\frac{1}{\sqrt{2\pi \sigma_i^2}}\exp\bigg[-\frac{(V_i-V_{h,i})^2}{2\sigma_i^2}\bigg];
\end{equation}
where $\vec{S}=(M,f_m,f_c,V_0,h,\Omega_\Lambda,\sigma_m)$ is a vector that encompasses the model parameters; $D$ and $V_h$ are heliocentric distances and velocities, respectively, and $(l,b)$ Galactocentric coordinates. For a given set of $\vec{S}$ we use Equation~(\ref{eq:kep}) to find the LG-centric velocity
 $v=v(r,M,t_0)$, which is then corrected by the apex motion~(\ref{eq:apex}) as $V_i=v(r_i,M,t_0) - {\bf v}_{\odot\r\lg} \cdot \hat {\bf r}_{i\r\odot}$ and inserted into the likelihood~(\ref{eq:likelv}).

Equation~(\ref{eq:kep}) is solved upon the assumption that galaxies in the LV move on radial orbits. Peculiar motions about the Hubble flow are accounted for by the hyperparameter $\sigma_m$, which we introduce in the two-dimensional variance of the $i$-th measurement as $\sigma_i^2=\epsilon_{V,i}^2 + \epsilon_{D,i}^2 (\d V/\d D)^2_{(D_i,l_i,b_i)} + \sigma_m^2.$ Here $\epsilon_{V}$ and $\epsilon_{D}$ denote, respectively, velocity and distance errors (see Ma et al. 2013), while $\d V/\d D$ corresponds to the velocity gradient at a given location $(D,l,b)$ returned by our model. Tests carried by P14 show that marginalizing over $\sigma_m$ yields unbiased joint bounds on the parameters of interest insofar as the tangential velocities that give rise to the peculiar motions are randomly oriented on the sky.


We use a Bayesian inference code, {\sc MultiNest} (Feroz \& Hobson 2008), to produce posterior samplings for our parameters and measure the error of the evidence. Flat priors are adopted for the parameters of interest ($M$, $f_m$, $f_c$, and $\sigma_m$) and Gaussian priors for parameters taken from the literature ($V_0$, $H_0$ and $\Omega_\Lambda$). 

The data consist of a catalogue $N_g=35$ systems compiled by P14 within barycentric distances $0.78 \lesssim r/\mpc \lesssim 3.0$ and known line-of-sight velocities. Most galaxies at $r\lesssim 0.78\mpc$ are satellites of either our Galaxy or M31 and move on orbits that are poorly described by Equation~(\ref{eq:kep}).
Beyond 3 Mpc three prominent associations introduce significant perturbations in the Hubble flow: IC 342/Maffei-I, M81 and Centaurus A/M83. To minimize such perturbations
P14 only include galaxies whose distance to any of those three major associations is larger than 1 Mpc.

In this work we also construct models that fit simultaneously the dynamics of galaxies in the LV {\it and} the relative motion of our Galaxy and M31. Following Lahav et al. (2000) we consider a joint likelihood weighted by the number of objects in each sample, $(N_g+1)\ln \mathcal{L}=N_g\ln \mathcal{L}_{\rm LV}+\ln \mathcal{L}_{\rm tim}$, where 
\begin{equation}\label{eq:likelg}
\ln \mathcal{L}_{\rm tim}(D,v_{\rm los},v_W,v_N|\vec{S})= \sum_{i=1}^4\bigg(-\frac{\ln[2\pi\sigma_i^2]}{2}-\frac{[X_i(\vec{S})-x_i]^2}{2\sigma_i^2}\bigg),
\end{equation}
and $x_i=\{D,v_{\rm los},v_W,v_N\}$ denote heliocentric measurements of M31 with an error $\sigma_i$. $X_i(\vec{S})$ correspond to the model prediction from Equations~(\ref{eq:rtkep}) and~(\ref{eq:vkep}) expressed in a heliocentric frame.

\section{Results}\label{sec:results}
Fig.~\ref{fig:grid} displays posterior distributions for each model parameter as returned by {\sc MultiNest}. 
Note first that the dynamics of galaxies in the LV (left panels) do not provide enough information to constrain the total mass of the LMC. Indeed, Fig.~\ref{fig:grid} shows a strong degeneracy between the parameters $f_c=M_\lmc/M_\mw$ and $f_m=(M_\lmc+M_\mw)/M_\and$. Adopting a LMC mass of $f_c\approx 0.01$ as in previous works yields models where Andromeda is approximately twice as massive as our Galaxy ($f_m\simeq 0.5$), thus recovering the main result of D14 and P14. Relaxing this assumption, however, weakly favours solutions where LMC is a factor $\sim 10$ more massive than suggested by its internal kinematics, i.e. $f_c\sim 0.2$, although models where $f_c=0.01$ cannot be ruled out with high significance. Our analysis also returns a constraint on the LG mass, $M\simeq 2\times 10^{12}M_\odot$, which appears to be insensitive to the mass ratios $f_c$ and $f_m$. Fig.~\ref{fig:timing} shows that this value is a factor $\sim 2.2$ lower than expected from the timing argument with no Magellanic Clouds ($\sim 4.5\times 10^{12}M_\odot$), and suggests that a massive LMC may help to ease this discrepancy.

Indeed, fitting the dynamics of nearby galaxies and the relative motion between our Galaxy and M31 {\it simultaneously} (right panels) provides strong support for a massive LMC. We find 
$M_\lmc=0.25_{-0.08(-0.15)}^{+0.09(+0.19)}\times 10^{12}M_\odot$ at 68\% (95\%) confidence, which rules out models with $M_\lmc\lesssim 0.1\times 10^{12}M_\odot$ at a 95\% confidence level. Such a large mass is indicative of an extended dark matter halo surrounding LMC and supports the scenario where this galaxy is currently undergoing its first pericentric passage around the MW. 

\begin{figure}  
\begin{center}
\includegraphics[width=80mm]{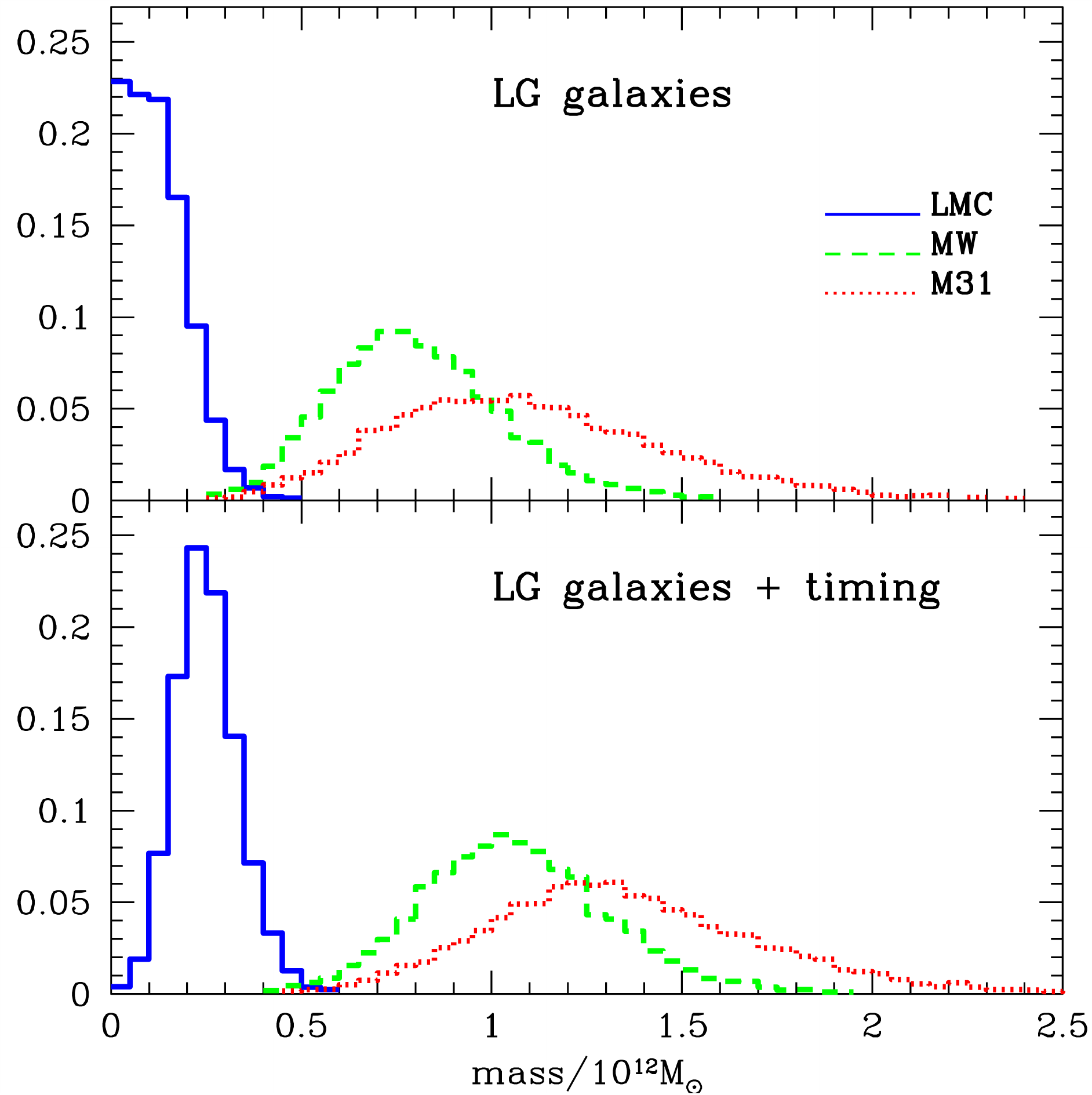}
\end{center}
\caption{Posterior distributions for the masses of the LMC, MW and M31. The upper panel shows posteriors obtained from the dynamics of galaxies in the LV. The lower panel fits simultaneously the dynamics of nearby galaxies {\it and} the relative motion between our Galaxy and M31.}
\label{fig:post_m}
\end{figure}

Fig.~\ref{fig:post_m} plots the posterior distributions for the individual masses of the LMC, Andromeda and the MW. Comparison between the upper and lower panels shows that the timing constraint shifts the mass of the three galaxies to slighlty higher values, $M_\mw=1.04_{-0.23(-0.44)}^{+0.26(+0.56)}\times 10^{12}M_\odot$, and Andromeda, $M_\and=1.33_{-0.33(-0.60)}^{+0.39(+0.88)}\times 10^{12}M_\odot$, while the LG mass increases to $M=2.64_{-0.38(-0.63)}^{+0.42(+0.91)}\times 10^{12}M_\odot$. Interestingly, in all our models M31 appears more massive than the MW, $M_\mw/M_\and\sim 0.75$, although the freedom in the value of $f_c$ adds uncertainty to this result.

\section{summary \& discussion}\label{sec:sum}

HST observations indicate that the LMC is slingshotting from the MW at a staggering $+327\kms$ (e.g. Kallivayalil et al. 2013). This implies that even a relatively light LMC may have a measurable impact on the Galaxy barycentre and the masses inferred from the timing argument.
Using the timing equations we simultaneously measure the (total) masses of the LMC, the MW and M31 by fitting the locations and radial velocities of a catalogue of 35 nearby galaxies ($\lesssim 3\mpc$) compiled by P14. An upper limit $M_\lmc\lesssim 0.31\times 10^{12}M_\odot$ is found with 95\% confidence. The LMC mass is further constrained from below by fitting the dynamics of nearby galaxies {\it and} the relative motion between the MW and Andromeda, which yields $M_\lmc=0.25_{-0.08(-0.15)}^{+0.09(+0.19)}\times 10^{12}M_\odot$ at a 68\% (95\%) confidence levels. Our method also puts bounds on the masses of the MW, $M_\mw=1.04_{-0.23(-0.44)}^{+0.26(+0.56)}\times 10^{12}M_\odot$, and Andromeda, $M_\and=1.33_{-0.33(-0.60)}^{+0.39(+0.88)}\times 10^{12}M_\odot$, which are consistent at one-sigma level with previous measurements that neglect the impact of the LMC on the observed Hubble flow (D14, P14). 
Our results indicate that LMC represents $\sim 1/5{\rm th}$ of the Galaxy mass, $M_\lmc/M_\g=(0.19\pm 0.05)\times 10^{12}M_\odot$. That is $\sim 10$ times higher than the mass enclosed within its nominal luminous radius, which suggests the existence of an extended dark matter halo surrounding this galaxy and supports the scenario where the LMC is currently undergoing its first pericentric interaction with the MW. 

In addition to the tests of P14 we have inspected two of our main assumptions: (i) galaxies can be modelled as point masses and (ii) Large Scale Structures (LSS) do not perturb the local Hubble flow ($<3\mpc$). Using live $N$-body models of G\'omez et al. (not shown here) we find that tidal stripping and dynamical friction only affect our timing mass estimates at a $\lesssim 5\%$ level. This result shows little sensitivity to the choice of density profile for the LMC.
Also, motivated by Libeskind et al. (2015) we have searched for signatures of a velocity shear in our galaxy sample arising from the motion of the LG through the cosmic web, but found no statistically-meaningful deviation between the velocities predicted by the point-mass model and the location of galaxies on the sky. The accurate description of the local Hubble flow by the timing argument suggests that neglecting the impact of the LSS on the dynamics of nearby galaxies does not unduly affect our constraints.


The results presented here call for theoretical and observational follow-up work. For example, live $N$-body experiments are needed to understand the perturbations that a massive LMC may induce on the Galaxy constituents (disc, bulge and halo). In addition, extending the work of Li \& White (2008) and Gonz\'alez et al. (2014) to the Local Volume will help to further test our method and calibrate the dynamical (e.g. virial) masses of the LMC, MW and M31 from the measured timing masses. Future kinematic surveys of the stellar populations discovered beyond the nominal luminous radius of the LMC (e.g. Majewski et al. 2009)
 will complement the mass estimate derived here using cosmological arguments.



\section*{Acknowledgements}
We thank Roeland van der Marel for very useful comments. The research leading to these results has received ERC funding under the programme (FP/2007-2013)/ERC Grant Agreement no. 308024.

{}

\end{document}